\documentclass[aps,twocolumn,showlabels,showrefs,amsmath,amssymb,pre,superscriptaddress, floatfix, colors]{revtex4-1}

\usepackage{graphicx}
\usepackage{dcolumn}
\usepackage{bm}
\usepackage{amssymb}
\usepackage{mathtools}
\usepackage{multirow}
\usepackage{bbold}

\newcommand{\1}{\begin{equation}}
\newcommand{\2}{\end{equation}}
\newcommand{\ea}{\begin{eqnarray}} 
\newcommand{\ee}{\end{eqnarray}}

\newcommand*{\bbar}[1]{\bar{\bar{#1}}}

\begin{document}
\title{Active particles in non-inertial frames: how to self-propel on a carousel}

\date{\today}

\pacs{..}

 \author{Hartmut L\"owen}
 \affiliation{Institut f\"{u}r Theoretische Physik II: Weiche Materie, Heinrich-Heine-Universit\"{a}t D\"{u}sseldorf, D-40225 D\"{u}sseldorf, Germany}


\begin{abstract}
Typically the motion of self-propelled active particles is described in a quiescent environment
establishing an inertial frame of reference.
Here we assume that friction, self-propulsion and fluctuations occur relative to a non-inertial frame
and thereby  the active Brownian motion model is generalized to non-inertial frames.
First, analytical solutions are
presented for the overdamped case, both for linear swimmers and circle swimmers.
For a frame rotating with constant angular velocity ("carousel"),
the resulting noise-free trajectories in the static laboratory frame trochoids if these are circles in the rotating frame.
For systems governed by inertia, such as vibrated granulates
or active complex plasmas, centrifugal and Coriolis forces become relevant. For both linear and circling
self-propulsion, these forces lead to out-spiraling trajectories which for long times approach a {\it spira mirabilis}.
This implies that a self-propelled particle will typically leave a rotating carousel.
A navigation strategy is proposed to avoid this expulsion, by adjusting the self-propulsion direction at wish.
For a particle, initially quiescent in the rotating frame, it is shown that this
strategy only works if the initial distance to the rotation centre is smaller than
a critical radius $R_c$ which scales with the self-propulsion velocity. Possible experiments to verify the theoretical predictions are discussed.
\end{abstract}
\maketitle

\section{Introduction}\label{ra_sec1}

One of the basic principles of classical mechanics is that Newton's second law holds only
in inertial frames of reference. If one transforms Newton's second law into a non-inertial frame,
there are additional fictitious forces such as the centrifugal force and the Coriolis force
which have to be added to describe the equations of motion in the non-inertial frame \cite{Goldstein}.
The coordinates of the particle trajectories can then be calculated either in the inertial frame or
in the non-inertial frame provided the additional fictitious (or inertial) forces are taken into account in the latter.

Recently active (or self-propelled) particles have
been studied intensely  by adding extra internal propulsion forces to the Brownian equations of motion of passive particles.
An active particle
possesses an intrinsic orientation along which it is self-propelling and therefore the
equations of motion involve both a force and a torque balance.
Clearly, as the autonomous motion of active particles needs a steady
conversion of intrinsic energy into mechanical motion, the motion is non-Hamiltonian and
describes a non-equilibrium phenomenon. Friction is typically involved and essential.

Examples for self-propelled particles
include  animals and
microorganisms as well as inanimate synthetic particles such as Janus colloids, dusty plasmas,
and vibrated granulates. In this flourishing realm of physics
\cite{Elgeti,Gompper2016,our_RMP,Ramaswamy2010,Menzel2015,Zottl2016,Guix}
most of the studies assume a quiescent plane of motion as a reference frame.
At low Reynolds number, i.e. in the limit when inertial effects can be neglected,
typical trajectories of self-propelled particles are linear ("a linear swimmer")
\cite{Houwse} or circular (a "circle swimmer") \cite{van_Teeffelen_Loewen_PRE_2008,Kuemmel_et_al_PRL_2013,Kurzthaler,Stocco}.
The latter type of swimmers experiences also a torque which steadily changes the direction of self-propulsion. These chiral swimmers were
studied in various environments
\cite{Marchesoni2,Lammert,Marchesoni1,Sevilla,Levis,Klapp,Franosch2}.

In this paper, we consider self-propelled particles in a non-inertial frame of reference such as a
rotating substrate ("carousel"). Even in the overdamped case, the swimmer trajectories are in general not
obtained by a simple coordinate transformation between the laboratory frame and the accelerated frame.
This is due to the fact that one has to specify which frictional forces are at work determining the dynamics:
this can be a friction proportional to the velocity in the accelerated frame, a friction
proportional to the velocity in the rest frame, or a combination of the two. The specified friction
will result in different equations of motion corresponding to different particle trajectories. Moreover
one has to define whether the self-propulsion occurs relative to the moving or rest frame,
and the same needs to be specified for the fluctuations (white noise). It is a bit surprising that - except for very recent work of active particles on a rotating spherical surface \cite{Dunkel} - 
this issue was not yet considered and explored in microswimmer physics;
what has been addressed is overdamped motion of swimmers
in an external flow field \cite{Reference_103_from_Kuechler,Kuechler,Mathijssen_rheotaxis,Golestanian}
but this flow field is typically different from that of a purely rotating fluid.

Here complete analytical solutions are
presented  both for linear swimmers and circle swimmers in case the white noise is the same in both frames.
For a frame rotating with constant angular velocity ("carousel"),
the resulting noise-free trajectories in the static laboratory frame  exhibit
{\it epicyclic swimming} with rosette-like trajectories \cite{sperm_spirograph}. The swimming paths
are epitrochoids or hypotrochoids if these are simple circles in the rotating frame.
Combinations of frictions are also considered and
can be mapped onto effective parameters of the analytical solutions.

Inertia in self-propelled systems is relevant for macroscopic self-propelled objects (such as
vibrated granulates, air planes, humans and animals) and to micron-sized dust particles in a plasma
(so-called complex plasma) \cite{Ivlev_Morfill}. The equations of motion are definitively different
in the rest and moving frame differing by the fictitious inertial forces as the centrifugal force and
the Coriolis force. Again specifying the kind of friction, self-propulsion and fluctuations relative to the two frames
will result in different classes of equations of motion. Still we always assume that self-propulsion
occurs in the non-inertial frame and that the same noise is acting in both frames.

Let us take one example to illustrate the situation:
consider a person running with constant speed on a turntable ("carousel"): the self-propulsion in
this case is dominated by interaction of the legs with the rotating ground, so it
is performed with a constant self-propulsion force in the moving frame. Moreover, there are two kinds of friction:
a friction relative to the moving ground which is proportional to the speed in the rotating frame and a friction with the
quiescent air  which is proportional to the speed in the rest frame. The actual type of trajectory
depends on the kind of friction specified.

With inertia and a constantly rotating disk, we show, for both linear and circling
self-propulsion, that the particle always performs trajectories which are spiraling outwards and
approach a logarithmic spiral ({\it spira mirabilis}) for long times. In other words,
the kinetic temperature of a many body system on a rotating disk increases exponentially for long times.
This is the same for passive and active particles.

Using the aforementioned example of a running person on a rotating disk,
everyday life experience indeed tells us that it is difficult to stay on the
rotating disk except if one is close to the rotation centre.
In fact, viewed in the rotating frame, the most dangerous force
driving the person outwards is the centrifugal force which is directed outwards away from the rotation centre.
Now one can ask a question of navigation strategy: Is it possible to run in a way to always stay on the disk?
This is particularly relevant if one can turn around quickly but does not change the translational speed.
We examine this question here and find indeed a navigation strategy via which a self-propelled object can
stay on the rotating disk forever if the direction of motion is  adjusted to the rotation centre. This is
obvious for the overdamped case but non-trivial for the underdamped case.
For  particles initially quiescent in the rotating frame, it is shown that the
strategy works if the initial distance to the rotation centre is smaller than
a critical radius
\begin{equation} \label{eq:0}
R_c= \gamma v_0/m\omega^2
\end{equation}
where $\gamma$ is the translational friction coefficient, $v_0$ is the self-propulsion speed in the rotating frame,
$m$ is the particle mass and $\omega$ the constant angular rotation velocity. The critical radius
can simply be understood as the threshold where the centrifugal force $mR_c \omega^2$ equals the
self-propulsion force $\gamma v_0$.

Corresponding noise-averages for the mean trajectories and the mean-square displacements in the
laboratory frame are also calculated and compared to that in the rotating frame. In particular,
analytical results are presented for linear accelerations and for overdamped
dynamics and constant rotation.

Our theoretical results can be tested in experiments. There are manifold realizations of self-propelled particles in
non-inertial reference frames, in particular for rotating frames.
Apart from the human motion on carousels \cite{carousel}, these range from  vibrated granulates on a vertically rotating
turntable  \cite{granulates}
(which belong also to the standard set-ups when horizontally rotated \cite{Iker,Briels})
birds and airplanes flying in the rotating atmosphere of the earth \cite{bird_Coriolis}, dust particles in a plasma
confined between rotating electrodes \cite{Bonitz,Hartmann1,Hartmann2}, as well as
beetles \cite{beetles}  and microswimmers
moving in a rotating fluid.

The paper is organized as follows: First, in section II, we discuss self-propulsion in a non-inertial
frame rotating with constant angular velocity. We discuss the overdamped case in detail and then include inertial effects.
A swimming strategy to stay close to the origin is also proposed. Then, in section III, we consider
translationally accelerated frames. Experimental realizations are discussed in section IV and we conclude in section V.

\section{Constant rotation}\label{ra_sec1}
\subsection{Overdamped case}\label{ra_subsec1}

We consider an inertial laboratory frame and a frame rotating with constant angular velocity ${\vec \omega}= (0, 0, \omega)^T $ with respect to the laboratory frame around their joint $z$-axis where the
superscript $^T$ just means a transposition of a row vector to a column vector
and $\omega>0$ is a rotation in the mathematical positive sense. Any time-dependent vector ${\vec a}(t) = (a_x(t), a_y(t), 0)^T$ in the $xy$-plane of the laboratory frame 
is then transformed into a corresponding vector ${\vec a}' (t)$ in the non-inertial rotating frame as mediated by the rotation matrix
\begin{equation} \label{eq:4}
\bbar{D}(\omega t) = 
\begin{pmatrix*}[c]
\cos{\omega t} & -\sin{\omega t} & 0 \\
\sin{\omega t} & \cos{\omega t} & 0 \\
0 & 0 & 1 \\
\end{pmatrix*}
\end{equation}
such that
\begin{equation} \label{eq:5}
\vec{a}(t) = \bbar{D}(\omega t) \vec{a}'(t)
\end{equation}
where we have assumed without loss of generality that the two frames coincide at time $t=0$. Clearly the rotation matrix fulfills
\begin{equation} \label{eq:7}
\bbar{D}^{-1}(\omega t) = \bbar D (-\omega t), \, \bbar D (\omega_1 t) \bbar D (\omega_2 t) = \bbar D (\omega_1 t + \omega_2 t)
\end{equation}
Time derivatives in the rotating and laboratory frame are denoted with $\text{d}/\text{d}t |'$ or with $\text{d}/\text{d}t$ respectively and are related via 
\begin{equation} \label{eq:7a}
\frac{\text{d}}{\text{d}t}\vec{a}'(t)\bigg\vert' = \dot{\vec a}' (t) - \vec\omega \times {\vec a}' (t)
\end{equation}
for any vector ${\vec a}'(t)$. 

Now we write down equations of motion for a single active Brownian particle in the inertial laboratory frame.  If the motion is confined to the $xy$-plane, the particle location is described by its position vector ${\vec r} (t) = (x(t), y(t), 0)^T$ and it
  self-propels along its unit orientation vector $\hat n(t) = (\cos \phi (t) , \sin \phi (t),0 )^T$
where $\phi (t)$ is the instantaneous orientation angle relative to a fixed axis in the laboratory frame. In our first most fundamental model, we assume that the damping of the particle motion is proportional to the relative velocity 
$\dot{\vec{r}} - {\vec \omega} \times {\vec r}$ as ${\vec \omega} \times {\vec r}$ is the velocity of the rotating frame viewed from the laboratory frame. Then, the fundamental equations of active Brownian motion read as follows:
\begin{equation} \label{eq:1}
\gamma (\dot{\vec{r}}-\vec{\omega} \times \vec{r}) = \gamma v_0 \hat{n} + \vec{f}(t)
\end{equation}
\begin{equation} \label{eq:2}
\gamma_{R}({\dot \phi}-\omega) = M + g(t)
\end{equation}
These equations couple the particle translation and rotation and represent a force and torque balance. 
In detail,  $\gamma$ denotes a translational friction coefficient,
$v_0$ is the self-propulsion speed of the active particle,
and the components of ${\vec f} (t)$ and $g(t)$ are Gaussian random numbers
with zero mean and variance representing white noise from the surrounding, i.e. $\overline{\vec{f}(t)}=0$, $\overline{f_i(t_1)f_j(t_2)}=2 k_B T \gamma \delta_{ij}\delta(t_1-t_2)$, $\overline{g(t)}=0$, $\overline{g(t_1)g(t_2)}=2 k_B T \gamma_R \delta(t_1-t_2)$ where the overbar means a noise average.
Here $k_BT$ denotes an effective thermal energy quantifying the noise strength.
Finally $\gamma_R$ is a rotational friction and only the relative angular velocity
$\dot{\phi} - \omega$ in the rotating frame is damped. The quantity $M$ is an external or internal torque
that leads to circular motion. For $M=0$ and $\omega =0$, the self-propelled motion is linear along the particle orientation
(a {\it linear swimmer}),
for nonvanishing $M$ there is a systematic rotation in the particle orientation leading to circular motion
(a {\it circle swimmer})
\cite{my_review}.

We now transform the equations of motion (\ref{eq:1})  and  (\ref{eq:2})  from the laboratory frame to the rotating frame by applying the rotation matrix (\ref{eq:4})  to (\ref{eq:1}). Using (\ref{eq:7a})  we obtain
\begin{equation} \label{eq:10}
\gamma  \frac{\text{d}}{\text{d}t}\vec{r}'\bigg\vert'  = \gamma v_0 \hat n ' + \vec f'(t)
\end{equation}
\begin{equation} \label{eq:11}
\gamma_ R \frac{\text{d}}{\text{d}t}\phi'\bigg\vert' = M' + g'(t)
\end{equation}
Here transformed quantities are denoted with a prime $'$, so we used the transformed vectors
\begin{equation} \label{eq:6}
 \begin{aligned}
\vec{r}(t) &= \bbar{D}(\omega t) \vec{r}'(t), \,\hat{n} = \bbar{D} (\omega t) \hat n'(t) \\
\vec f(t) &= \bbar D (\omega t) \vec f' (t)
 \end{aligned}
\end{equation}
Since the Gaussian white noise is isotropic, it is clear that $\vec f'$ has the same statistics as $\vec f$. Hence noise averages are the same for $\vec f'$ and $\vec f$ and therefore $\vec f'$ and $\vec f$ can be identified. Moreover, $g'(t)=g(t)$, and  the torque is not affected by the transformation into the body frame, hence $M'=M$. Finally, the transformed orientation $\phi'=\phi-\omega t$ is the angle of $\hat n'(t)$ relative to a fixed axis in the rotating frame.

Consequently, the equations of motion in the body frame are that of an ordinary Brownian circle swimmer
\cite{van_Teeffelen_Loewen_PRE_2008}. Therefore we can adopt the solution for a circle swimmer in the rotating frame and transform it back to the laboratory frame via
$\vec{r}(t) = \bbar{D}(\omega t) \vec{r}'(t)$.

\subsubsection{Noise-free limit}

In the case of vanishing noise (${\vec f}(t)=0$, $g(t)=0$) the trajectories are deterministic.
In the rotating frame, (\ref{eq:10}) and (\ref{eq:11}) have the solutions of a circular trajectory with  a spinning frequency

\begin{equation} \label{eq:12}
\omega_s = \frac{M}{\gamma_R}
\end{equation}
and a spinning radius

\begin{equation} \label{eq:13}
R_s = \frac{ v_0 \gamma_R}{M}
\end{equation}
In detail, the solutions in the rotating frame are given by

\begin{equation} \label{eq:14}
 \begin{aligned}
\phi' (t)& = \phi'(0) + \omega_s t \\
\vec{r}' (t) &= \vec{r}'(0) + v_0 \int_0^t \text{d}t' \bbar D (\omega_s t') \hat n' (0) \\
&= \vec{r}' (0) + \frac{v_0 \gamma_R}{M}
\begin{pmatrix*}[c]
-\sin{\omega_s t}& 1-\cos{\omega_s t} \\
-1+\cos{\omega_s t} & -\sin{\omega_s t} \\
\end{pmatrix*}
\hat n ' (0) \\
&=\vec{R}_m + R_s \bbar{D}(\omega_s t - \frac{\pi}{2}) n'(0)
 \end{aligned}
\end{equation}
which describes a circle of radius $R_s$ centered around 
\begin{equation} \label{eq:15a}
\vec{R}_m= \vec{r}'(0) - R_s 
\begin{pmatrix*}[c]
0 & 1 \\
-1 & 0 \\
\end{pmatrix*}
\hat n'(0)
\end{equation}
Conversely, in the laboratory frame, the solutions are gained as
$\phi (t) = \phi ' (t) + \omega t$ and  $\vec r (t) = \bbar D (\omega t) \vec{r}' (t)$
such that we obtain
\begin{equation} \label{eq:17a}
\vec{r}(t) = \bbar{D}(\omega t) \vec{R}_m + R_s \bbar{D}((\omega_s+\omega)t-\frac{\pi}{2}) \hat{n}(0)
\end{equation}
using (\ref{eq:7}) and $\hat{n}'(0) = \hat{n}(0)$. Equation (\ref{eq:17a}) therefore has the mathematical interpretation that the circular trajectory $\vec{r}' (t) $  in the rotating frame
is transformed to an {\it  epicycle} in the  laboratory frame, a circle whose centre moves round the circumference of a another circle, i.e.\
the full trajectory is a superposition of two circular ones with two different radii $R_m$ and $R_s$
and two angular frequencies $\omega$ and $\omega_s$. 
Examples for these rosette-like  trajectories resulting from (\ref{eq:17a}) are displayed in Figure 1 including different special cases. In Figure 1a, the simplest special case
of a linear swimmer, $\omega_s=0$, is shown. The linear trajectory in the rotating frame
transforms into a degenerated epicycle in the laboratory frame.
The next special case plotted in
Figure 1b, is $\omega_s=-\omega$, i.e. when the two circular motions are exactly counterrotating. Then Eqn.\ (\ref{eq:17a}) implies that the resulting trajectory in the laboratory frame is a simple circle but with a shifted centre. 
For $\omega_s/\omega>-1$ the trajectories are {\it epitrochoids\/} \cite{trochoids}.
Is $\omega_s/\omega$ rational, the trajectories are closed, see Figure 1c, while they cover
the full ring area for an irrational ratio $\omega_s/\omega$, see Figure 1d.
Finally, for $\omega_s/\omega<-1$ the trajectories are {\it hypotrochoids\/} \cite{trochoids}, an example for a rational ratio is provided in Figure 1e.
We remark three points here: i) In a frame rotating with $-(\omega_s + \omega)$ relative to the laboratory frame, the trajectories are also simple circles and the self-propulsion force is constant in this frame. ii) The solutions (\ref{eq:17a}) are marginally stable upon a change in the initial conditions. iii) Similar rosette-like trajectories have been also found for swimmers in external potentials \cite{ten_Hagen_Nature_Comm_2014,Grier,Bertin} and for sperm on substrates \cite{sperm_spirograph}.

 \begin{figure}
  \centering
  \includegraphics[width=0.45\textwidth]{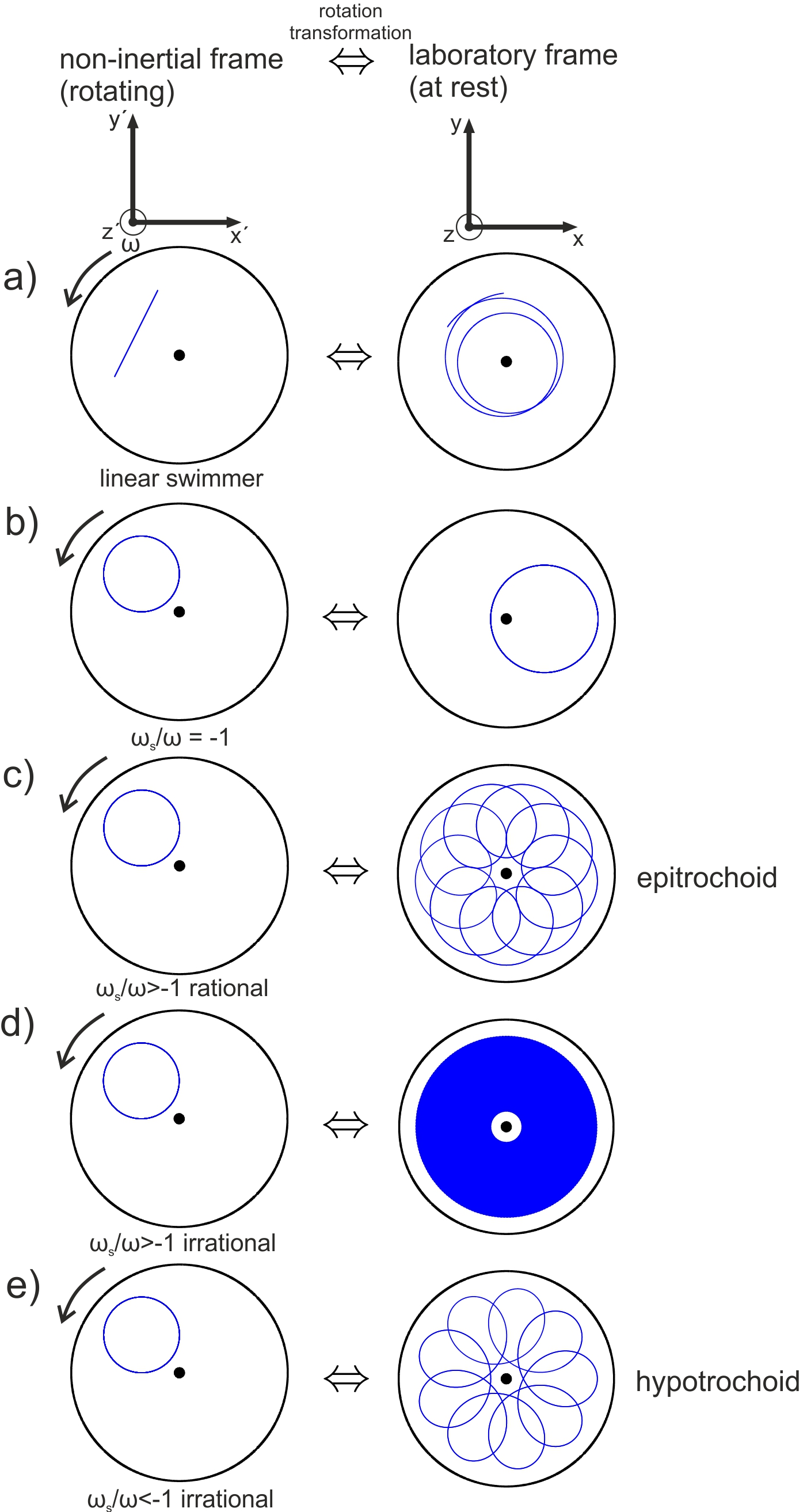}
  \caption{ Five examples of circular motion in the rotating frame (left column) and the mapping back to
the static laboratory frame (right column) via the rotation matrix indicated by the double arrow. The
static origin of the rotation is marked as a bullet point. a) degenerate case of a linear swimmer $\omega_s=0$. b) special case $\omega_s/\omega=-1$ leading to circular trajectories in both frames. c) rational ratio $\omega_s/\omega>-1$ leading to epitrochoids.
d) incommensurate frequencies when $\omega_s/\omega$ is irrational leads to covering of a ring-like area in the laboratory frame. e) rational ratio $\omega_s/\omega<-1$ leading to hypotrochoids.}
\label{Fig1}
\end{figure}

\subsubsection{Effects of noise}

We now address the noise averaged displacement for a prescribed initial orientation ${\hat n}(0)$ at time $t=0$.
Let us first discuss this quantity  in the rotating body frame. It is given by
$\overline{ {\vec r}'(t) - {\vec r}'(0)}$ where $\overline{\vphantom{K}...}$ indicates a noise-average.
The quantity  $\overline{ {\vec r}'(t) - {\vec r}'(0)}$ 
measures the mean displacement a self-propelled particle has achieved after a time $t$ provided its orientation
${\hat n}'(0)$ has been prescribed at time $t=0$. According to  the equations
(\ref{eq:10}) and (\ref{eq:11}), we can use the standard results for a linear swimmer ($M=0$) and a circle swimmer.
For a linear swimmer \cite{Houwse,tenHagen_JPCM}

\begin{equation} \label{eq:18}
\overline{\vec{r}'(t)-\vec{r}'(0)} = \frac{v_0}{D_r} (1-e^{-D_r t}) \hat n' (0)
\end{equation}
This represents a linear segment oriented
along ${\hat n}' (0)$ whose total length is the persistence length  $\ell_p = v_0/D_r$ of the
random walk and $D_r=k_BT/\gamma_R$ denotes the rotational diffusion constant.
In the general case of a circle swimmer, it is a logarithmic spiral ({\it "spira mirabilis"}) given by \cite{van_Teeffelen_Loewen_PRE_2008}:
\begin{equation} \label{eq:19}
 \begin{aligned}
\overline{\vec{r}'(t)-\vec{r}'(0)} &= \lambda (D_r \hat n'(0) + \omega_s \hat n'^{\perp}(0)  \\
&- e^{-D_r t} (D_r \bar{\hat n}' + \omega_s \bar{\hat n}'^{\perp}))
 \end{aligned}
\end{equation}

\begin{equation*} 
 \begin{aligned}
\text{with} \; \lambda &= v_0/(D_r^2 + \omega_s^2) \\
\hat n'^{\perp}(0) &= (-\sin{\phi' (0)}, \cos{\phi'(0)}, 0)^T \\
\bar{\hat n}' &=(\cos{(\omega_s t+ \phi'(0))}, \sin{(\omega_s t +\phi'(0))}, 0)^T, \;\text{and}\\
\bar{\hat n}'^{\perp}& =(-\sin{(\omega_s t + \phi'(0))}, \cos{(\omega_s t + \phi'(0))}, 0) ^T
 \end{aligned}
\end{equation*} 

The spread of the mean displacement is embodied in the mean squared displacement (MSD) for which the general result
is known for circle swimmers \cite{van_Teeffelen_Loewen_PRE_2008}:
\begin{equation} \label{eq:20}
 \begin{aligned}
\overline{(\vec{r}'(t)-\vec{r}'(0))^2} &= 2 \lambda^2  \omega_s^2 - D_r^2 + D_r (D_r^2+\omega_s^2)t+ e^{-D_r t} \\
&\cdot [(D_r^2-\omega_s^2)\cos{\omega_s t}-2D_r \omega_s \sin{\omega_s t}] + 4Dt
 \end{aligned}
\end{equation}
where $D=k_BT/\gamma$ is the translational short-time diffusion constant.

We now calculate the mean displacement $\overline{ \vec{r}(t) - \vec{r}(0)}$ and the MSD in the laboratory  frame. In fact,
according to Eq.\ (\ref{eq:6}),
\begin{equation} \label{eq:21}
\overline{ {\vec r}(t) - {\vec r}(0)} 
=\bbar{D}(\omega t) (\overline{\vec{r}'(t)-\vec{r}'(0)}) + (\bbar{D}(\omega t)-\bbar{\mathbb{1}})\vec{r}(0) 
\end{equation}
Here $\bbar{\mathbb{1}} \equiv \bbar{D}(0)$ denotes the unit tensor.
Inserting the previous result (\ref{eq:19}) into Eq.\ (\ref{eq:21}) an explicit expression is gained
for the mean displacement in the laboratory frame. As is evident from Eq.\ (\ref{eq:21}), one part of the mean displacement is the
transformed previous one, another stems from the fact that the initial starting point ${\vec r}(t=0) ={\vec r}'(t=0)$
is fixed in the body system which gives rise to a rotated reference point also contributing to the displacement.
Results from the explicit expression for the mean trajectory are shown in Figure 2 for both a linear
swimmer and a circle swimmer. As a reference, the mean trajectories  (based on the expressions (\ref{eq:18}) and (\ref{eq:19})) are also
shown in the left column of Figure 2 in the rotating frame. The mean trajectories in the laboratory frame are assuming a quite complex shape due to the superposition of self-propulsion and rotation, in particular for the transformed {\it spira mirabilis} (Figure 2b).

 \begin{figure}
  \centering
  \includegraphics[width=0.45\textwidth]{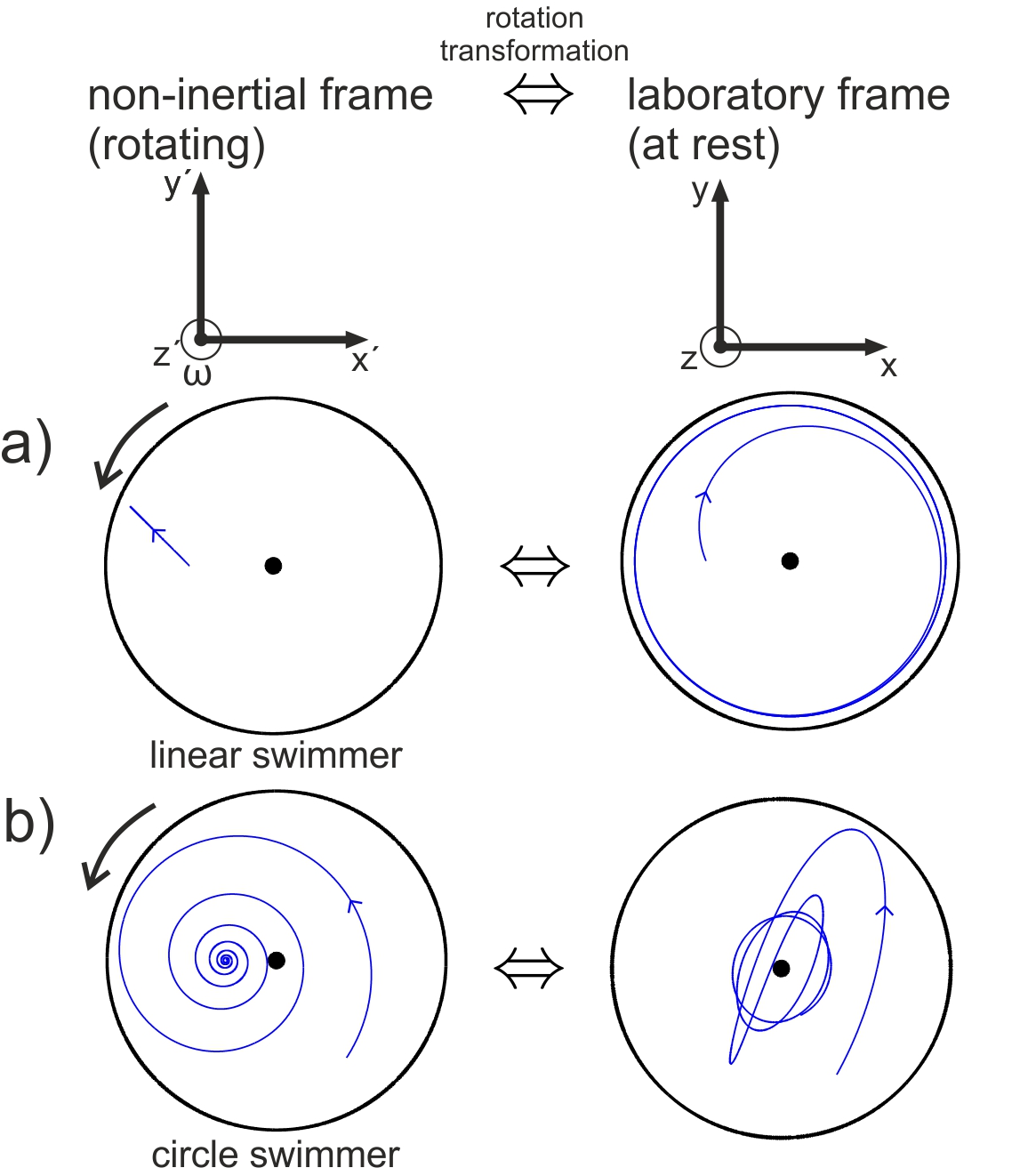}
  \caption{Same as Figure \ref{Fig1}, but now for the mean trajectory $\overline{\vec{r}'(t)}$ in the rotating frame (left column) and for the mean trajectory $\overline{ {\vec r}(t)}$ in the laboratory frame (right column): a) for a linear swimmer where the mean
trajectory in the rotating frame is a linear segment which is correspondingly distorted due
to the rotation. b) for a circle swimmer
where the mean trajectory in the rotating frame is a logarithmic spiral which is again transformed
in the laboratory system.}
\label{Fig2}
\end{figure}

The MSD in the laboratory frame is obtained via
\begin{equation} \label{eq:22}
\begin{aligned}
&\overline{(\vec{r}(t)-\vec{r}(0))^2}  =\overline{(\vec{r}'(t)-\vec{r}'(0))^2} \\
&+ 2\left((\bbar D (\omega t)-\mathbb{1})\vec{r}(0)\right)\cdot\left(\bbar D (\omega t) \overline{(\vec{r}'(t)-\vec{r}'(0))}\right) \\
&+ \left((\bbar D (\omega t)-\mathbb{1})\vec{r}(0)\right)^2 
\end{aligned}
\end{equation}
which is again an explicit expression when Eqns.\ (\ref{eq:19}) and (\ref{eq:20}) are inserted here. A comparison between the MSDs
in the rotating and lab frame is provided in Figure 3. Again the co-rotating initial reference point $\vec{r}(0)$
creates the difference between the two MSDs, obviously they coincide when 
$\vec{r}(0)=0$. 

Defining long-time translation diffusion coefficients
$D_L$ and $D_L'$ according to
Einstein's formula
\begin{equation} \label{eq:23}
 \begin{aligned}
D_L' &= \lim_{t \to \infty} \frac{1}{4t} \overline{(\vec{r}'(t)-\vec{r}'(0))^2}, \\
D_L &= \lim_{t \to \infty} \frac{1}{4t} \overline{(\vec{r}(t)-\vec{r}(0))^2},
 \end{aligned}
\end{equation}
we immediately recognize that $D_L=D_L'$, i.e., the rotation does not lead to a change in long-time diffusion.

 \begin{figure}
  \centering
  \includegraphics[width=0.45\textwidth]{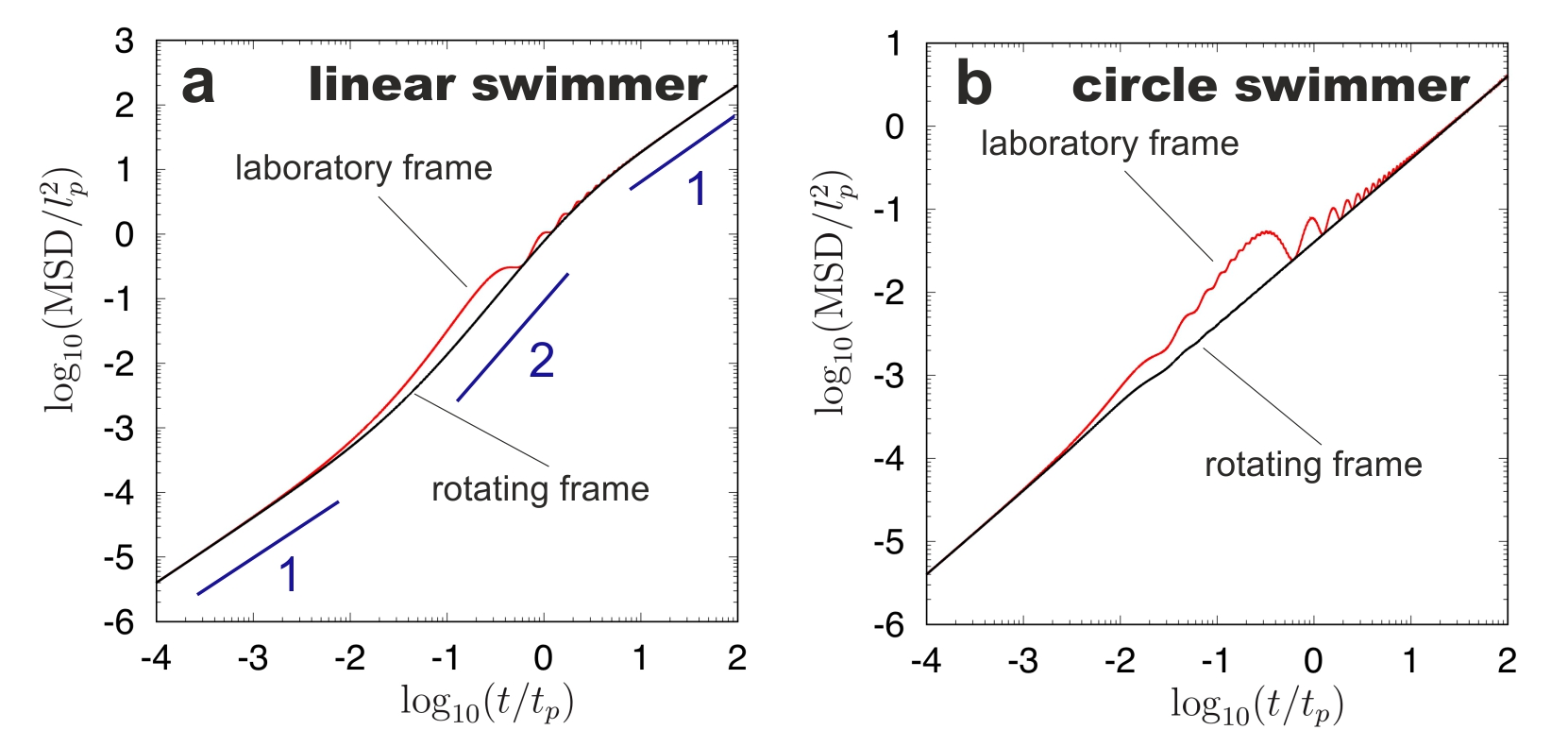}
  \caption{Double logarithmic plot of the mean-square displacement (MSD) as a function of time $t$.
a) for linear swimmers, b) for circle swimmers.  Length and time units
are the persistence length $\ell_p = v_0/D_r$ and persistence time $t_p = 1/D_r$. Energy units are in $k_BT$.
In these units, the further parameters are: a) $M=0$, $\gamma = 1$, $\gamma_R =0.1 $, $\omega = 1$. b)
$M= 2$, $\gamma =1 $, $\gamma_R = 0.1$, $\omega = 1$. Reference slopes of 1 and 2 are also indicated.}
\label{Fig3}
\end{figure}

\subsubsection{ Friction and self-propulsion relative to  both inertial and non-inertial frames}

We first comment on the situation when the translational and rotational friction
is solely proportional to the velocities in the laboratory frame. In this case the basic equations of motion read as
\begin{equation} \label{eq:X1}
 \begin{aligned}
\gamma \dot{\vec{r}} &= \gamma v_0 \hat n + \vec{f}(t) \\
\gamma_R \dot{\phi} &= M + g(t)
 \end{aligned}
\end{equation}
Clearly here the situation is reversed: There is circular swimming in the laboratory frame and the transformation
to the rotating frame will lead to epicyclic trajectories with corresponding results for the noise-averages.
So the role of the two frames in Figure 1 is interchanged.

In general, as mentioned in the introduction, there are cases where the 
friction depends on both the rotating frame and the laboratory frame velocities
(and angular velocities).
Assuming that the torque and the noise is the same in both frames, the equations of motion
now read as
\begin{equation} \label{eq:X2}
 \begin{aligned}
\gamma_1 \dot{\vec{r}} + \gamma_2(\dot{\vec r} - \vec{\omega}_2 \times \vec{r}) &= (\gamma_1+\gamma_2)  v_{02} \hat n + \vec{f}_2(t) \\
\gamma_{R1}  \dot{\phi} +   \gamma_{R2} (\dot{\phi}- \omega_2) &= M_2 + g_2(t)
 \end{aligned}
\end{equation}
with two translational friction coefficients $\gamma_1$ and $\gamma_2$ and two rotational friction coefficients $\gamma_{R1}$ and $\gamma_{R2}$. In order to
indicate that we consider a case of double friction now, we have introduced a notation with a subscript $2$ for angular frequency, self-propulsion speed, external torque and noises. A closer analysis of the equations of motion 
(\ref{eq:X2}) reveals that they can be mapped exactly on the original equations of motion
(\ref{eq:1}) and (\ref{eq:2}) provided the following parameter identification is performed:
$\gamma=\gamma_1+\gamma_2$, $\omega=\gamma_2 \omega_2/(\gamma_1 + \gamma_2)$, $v_0= v_{02}$, $\vec{f}(t) =\vec{f}_2(t)$, $\gamma_R= \gamma_{R1}+\gamma_{R2}$, $M=M_2+\gamma_{R2}\omega_2$, $g(t)=g_2(t)$.
Hence the physics does not change albeit the parameters need to be renormalized with respect to our basic original equation.

\subsection{Effects of inertia}\label{ra_sec1}

We now generalize the equations of motion including inertia. In the rest frame, we include a mass and an acceleration term  (see e.g. \cite{Baskaran2,Enculescu_Stark_PRL,Brady,Zippelius,Ldov,Shankar,Puglisi,Dauchot_hexbugs,Das}) as
\begin{equation} \label{eq:24}
 \begin{aligned}
m\ddot{\vec{r}} + \gamma(\dot{\vec r}-\vec{\omega}\times \vec r) &= \gamma v_0 \hat n + \vec{f} (t)\\
J \ddot\phi + \gamma_R (\dot\phi - \omega) &= M+g(t)
 \end{aligned}
\end{equation}
In many practical relevant cases the orientational relaxation is fast, hence the case of a vanishing moment of inertia,
$J=0$, is considered
subsequently as also done elsewhere \cite{Enculescu_Stark_PRL,Brady,Zippelius,Das}.


Again let us discuss the noise-free case first. For $J=0$, the solution of the equation of motion (\ref{eq:24}) is $\phi (t)= \phi(0)+ \omega_0 t$ with $\omega_0 = \omega + \omega_s$  and
\begin{equation} \label{eq:26}
\vec{r}(t)=\sum_{j=1}^2\left(C_j  \left( \begin{array}{c} 1 \\ i \\ 0 \end{array}\right) e^{\lambda_j t}+ \text{c.c.}\right) + \vec r_p (t)
\end{equation}
where the complex coefficients $C_1$ and $C_2$ can be determined in terms of the initial conditions for both
${\vec r}(0)$ and $\dot {\vec r}(0)$ and c.c. means complex conjugation. The complex eigenfrequencies
$\lambda_1$ and $\lambda_2$ from the homogeneous equation are given as
\begin{equation} \label{eq:27}
 \begin{aligned}
\lambda_1 &= -\frac{\gamma}{2m} (1+\Delta) + \frac{i \omega}{\Delta} \\
\lambda_2 &= \frac{\gamma}{2m} (\Delta-1) + \frac{i \omega}{\Delta} \\
\text{where}\; \Delta&=\sqrt{\frac{1}{2} + \sqrt{\frac{1}{4} + \frac{4 \omega^2 m^2}{\gamma^2}}} \; > \;1
 \end{aligned}
\end{equation}
Remarkably, the real part of $\lambda_1$ is negative while the real part of $\lambda_2$ is positive.
Hence the long-time dynamics will be dominated by $\lambda_2$.

\begin{figure}
  \centering
  \includegraphics[width=0.3\textwidth]{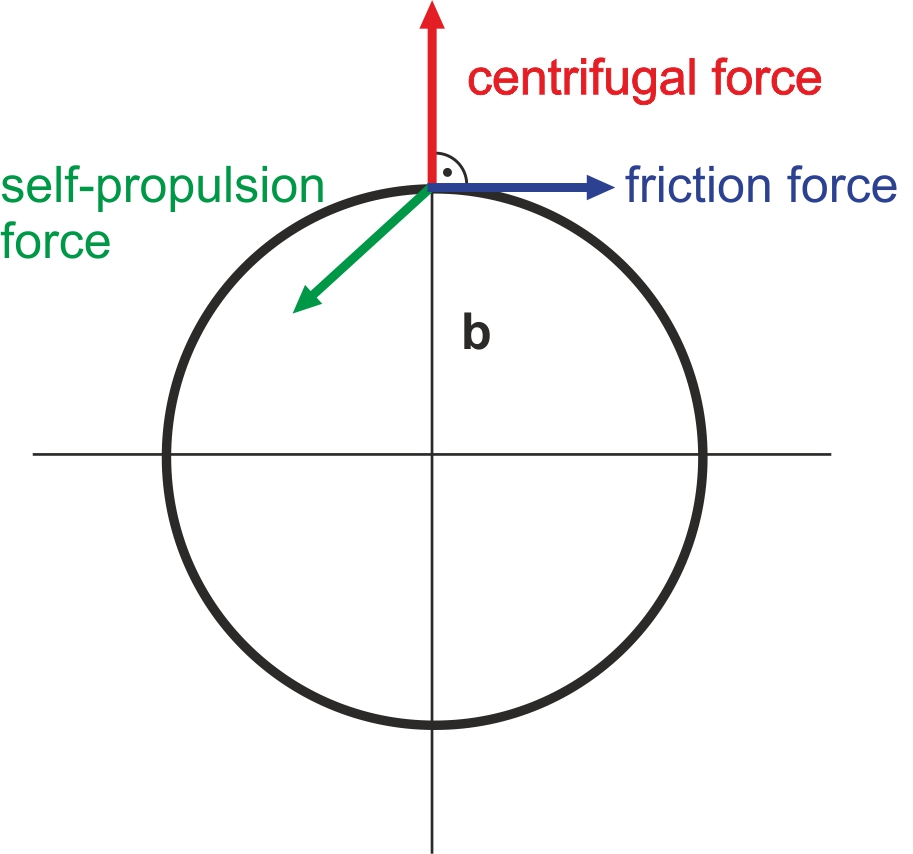}
  \caption{Static balance of the centrifugal force, the self-propulsion force and the friction force
in a frame co-rotating with angular velocity $\omega+\omega_s$ where all
forces are time-independent.}
\label{Fig5}
\end{figure}

A particular solution ${\vec r}_p(t)$ for the inhomogeneous equation
can be found as 
\begin{equation} \label{eq:28}
\vec{r}_p(t) = \text{Re}(\vec b e^{i \omega_0 t})
\end{equation}
with a complex vector $\vec b$ determined as
\begin{equation} \label{eq:29}
\vec b = \frac{\gamma v_0}{\sqrt{m^2 \omega_0^4 + \gamma^2\omega_s^2}} \left( \begin{array}{c} 1 \\ -i \\ 0 \end{array}\right)
\end{equation}
The particular solution is best explained in a rotating frame rotating with angular speed $\omega_0$  for which we use a double prime notation $''$ and the notation $\frac{\text{d}}{\text{d}t}\bigg\vert''$ means a time derivative in this frame. The equations of motion read in the noise-free case as
\begin{equation} \label{eq:31a}
 \begin{aligned}
&m \frac{\text{d}^2}{\text{d}t^2}\vec{r}''\bigg\vert'' + 2m \vec{\omega}_0 \times \frac{\text{d}}{\text{d}t}\vec{r}''\bigg\vert'' + m \vec{\omega}_0 \times (\vec{\omega}_0 \times \vec{r}'') \\
&+ \gamma \left( \frac{\text{d}}{\text{d}t}\vec{r}''\bigg\vert''  - \vec{\omega}_s\times\vec{r}''\right) = \gamma v_0 \hat n'' \quad \\&\text{with} \quad \frac{\text{d}}{\text{d}t}\phi''\bigg\vert''   = 0 \\
 \end{aligned}
\end{equation}
and a possible solution has a constant ${\hat n}''$ and ${\vec{r}}''$ directly providing
the balance condition
of self-propulsion force, centrifugal force and friction force, see Figure \ref{Fig5},
 $ (m\omega_0^2b)^2 + \gamma \omega_s ^2b^2 = (\gamma v_0)^2$ which yields the radius 
\begin{equation} \label{eq:30}
b = \frac{\gamma v_0}{\sqrt{m^2 \omega_0^4 + \gamma^2 \omega_s^2}} 
\end{equation}
To summarize
there are three contributions to the general solution (\ref{eq:26}). The first term associated with the eigenfrequency
$\lambda_1$ leads to an exponentially damped contribution which becomes irrelevant for long times.
This term describes a {\it spira mirabilis} spiraling inwards towards the rotation center.
The second term with the eigenfrequency
$\lambda_1$ leads to an exponentially exploding contribution which becomes dominant for long times.
This is likewise  a logarithmic spiral which is now spiraling outwards. The third contribution, i.e.
the particular solution $\vec{r}_p(t)$, describes a
circular motion around the origin with angular velocity $\omega_0$ and radius $b$.
This contribution is clearly bounded, depends on the self-propulsion speed $v_0$ but is unstable.

\begin{figure}
  \centering
  \includegraphics[width=0.5\textwidth]{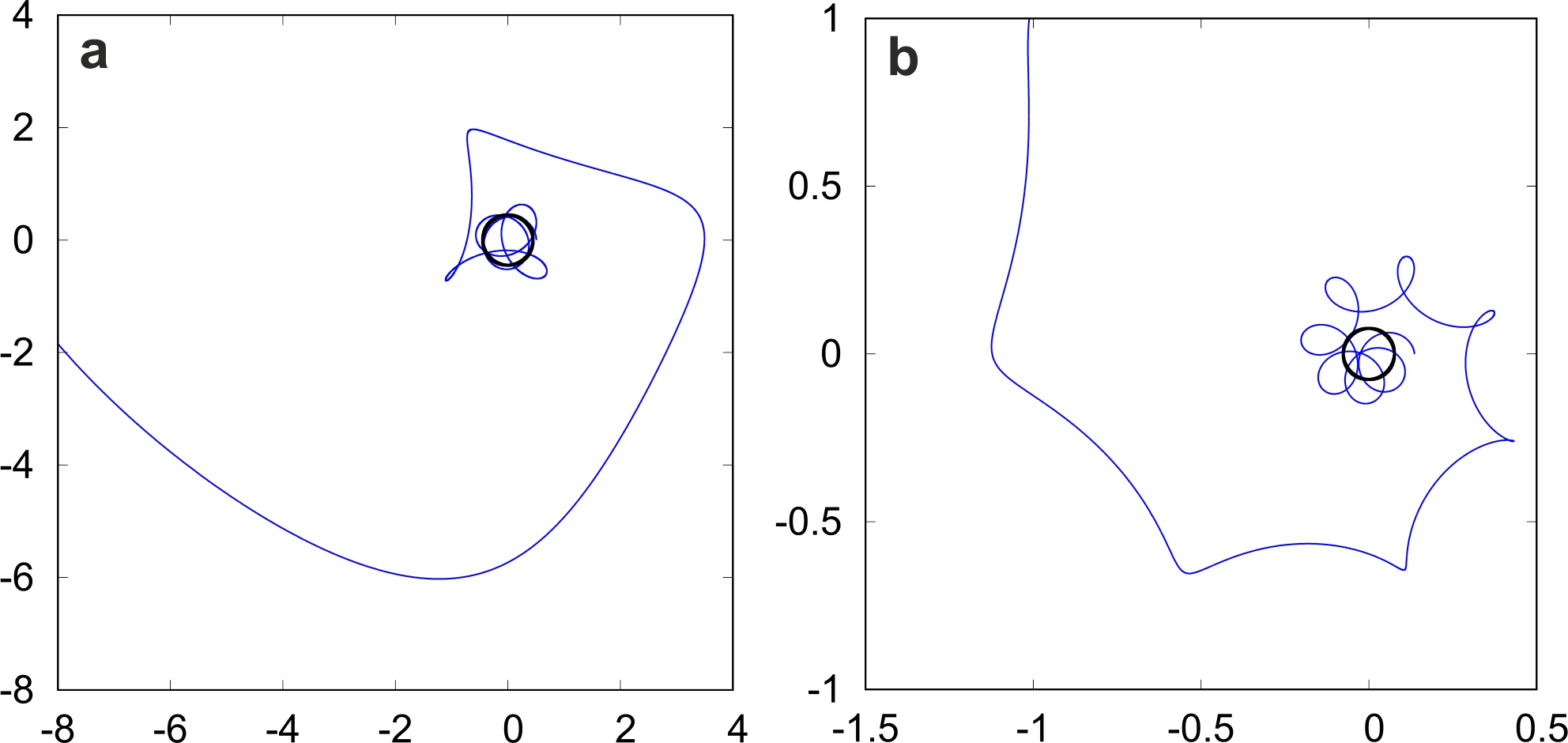}
  \caption{Typical trajectories from the analytical solution of Eq. (\ref{eq:26}) in the laboratory frame.
The 
unstable particular solution with radius $b$ is indicated as a black circle. The trajectory approaches a logarithmic spiral. 
The length unit is $v_0/\omega$ and
the parameters are: $\gamma / m\omega = 2$, $\gamma=0.1$, $m=0.5$ and a) $\omega_0/\omega = 2$ and b) $\omega_0/\omega = 5$}
\label{Fig4}
\end{figure}

It is important to note that any initial condition will lead to a spiral which exponentially grows in time
and therefore necessarily will leave any finite-sized turntable. The self-propelled particle will never
be able to stay inside a finite domain around the origin. Self-propulsion does not change this asymptotics,
the growing {\it spira mirabilis} generically occurs also for passive systems.
 Two examples with self-propulsion are displayed in Figure \ref{Fig4}.


If the radial distance of the particle from the origin
is large enough, the centrifugal force will dominate pushing away the particle from the origin.
Even a self-polarization strategy of the particle will not beat the centrifugal force at large distance,
the system is always unstable.
One way to obtain confinement of the particle to the origin of the rotation is a harmonic confinement
\cite{Szamel_PRE,Pototsky,volpe2013,nourhani2015,ribeiro2016,Soudeh} provided the strength
of the harmonic trap is larger than $m\omega_0^2$.

Finally we remark that the calculation of the noise averaged mean trajectory and particle MSD is much more complicated than in the overdamped case since these quantities depend not only on the initial orientation and particle location but also on the initial particle velocity. Moreover a direct mapping on the swimmer in the rest frame does not exist
as inertia will generate additional terms such as the centrifugal and Coriolis forces.  We leave this for future studies.
For more complicated friction terms as e.g.\ given by a superposition as in Eq.\ (\ref{eq:X2}), again a parameter mapping can be performed similar to what was discussed in the previous section.

\subsection{Navigation strategy}

Since the centrifugal force will drive a particle away from the origin, it is
interesting to explore under which conditions a noise-free self-propelled particle can reach the
rotation centre (or any other arbitrary target point) provided it
started with a vanishing initial velocity. We assume that the orientation of the particle can be
completely steered internally or from the outside, i.e. the swimmer has complete control over its swimming
direction but the swimming speed is fixed by $v_0$. This is a reasonable approximation for many animals
and humans and also for airplanes and spherical Janus-particles \cite{Cichos0,Cichos1,Cichos2,Cichos3,Grier,Liebchen_navigation}.

Let us first discuss the overdamped case where the navigation strategy turns out to be straight forward. Having the free choice of ${\hat n}'(t)$ in mind, one sees directly from Eqn.\ (\ref{eq:10})
that any target point ${\vec r}'_B$ can be
reached from any initial point ${\vec r}'_A$ by simply adjusting the self-propulsion orientation towards the
difference vector, ${\hat n}'= ({\vec r}'_B - {\vec r}'_A)/|  ({\vec r}'_B - {\vec r}'_A) |$.
Under this navigation strategy, the travel time from
${\vec r}'_A$ to ${\vec r}'_B$ is simply $|  ({\vec r}'_B - {\vec r}'_A) |/v_0$.
Due to the presence of centrifugal and Coriolis forces this simple argument does not hold any longer for the underdamped case. However, one can reach any inner target point
${\vec r}'_B$ from any  initial point ${\vec r}'_A$ (for $|{\vec r}'_B|<|{\vec r}'_A|$) if the condition
\begin{equation} \label{eq:100}
|{\vec r}'_A |< R_c= \gamma v_0/m\omega^2
\end{equation}
is fulfilled. We shall give a physical argument for that by proposing a strategy of multiple 
reversals  of the swimming direction: Let us start at rest at an initial position ${\vec r}'_A$ fulfilling $|{\vec r}'_A |< R_c$. Then the Coriolis force is zero and the centrifugal force can be overcome by the self-propulsion force in magnitude. Consequently, at time $t=0$,  the self-propulsion direction of the particle ${\hat n}'(0)$ can be chosen in such a way that the sum of the centrifugal and self-propulsion force is acting along ${\vec r}'_B - {\vec r}'_A$ towards the target point, i.e. the particle will be accelerated  towards the target. After a short time $\delta t$, the strategy is to reverse the particle orientation 
to ${\hat n}'(\delta t)=- {\hat n}'(0)$. Then, to linear order in $\delta t$, the particle stops moving at a time $2\delta t$ after having travelled a distance $\delta s$  towards the target. Repeating this procedure of particle orientation reversal, the particle is kept at small speeds but moves slowly towards the target. 
Summing over all displacements the particles finally arrives at the target. 

The navigation strategy of multiple reversals  works in particular when the target is the rotation centre, ${\vec r}'_B=0$, but it is not the fastest option to arrive at the origin.
Along a radial path, the best navigation strategy 
 is to choose the particle orientation to compensate
the tangential (polar) part of the Coriolis force (i.e. the projection of the
Coriolis force perpendicular to the motion) and use the remaining self-propulsion force to accelerate the particle towards the rotation centre.
In order to explore this in more detail, we rewrite the equations of motion
in polar coordinates in the rotating frame such that

\begin{equation} \label{eq:N1}
\vec{r}'=r' \hat e_r', \;\hat e_r' = \left( \begin{array}{c} \cos{\theta'} \\ \sin{\theta'} \end{array}\right), \; \hat e_{\theta}'=\left( \begin{array}{c} -\sin{\theta'} \\ \cos{\theta'} \end{array}\right)
\end{equation}
where $\theta'$ is the polar angle belonging to the position of the particle in the rotating frame.
Expressed in polar coordinates of the rotating frame, the equations of motion (\ref{eq:31a}) split into radial and polar parts and read as

\begin{align} 
m \frac{\text{d}^2}{\text{d}t^2}r'\bigg\vert' -mr' \frac{\text{d}}{\text{d}t}\theta^2\bigg\vert'  - mr'\omega^2 + \gamma \frac{\text{d}}{\text{d}t}r'\bigg\vert' - 2m \omega \frac{\text{d}}{\text{d}t}\theta'\bigg\vert' r' \notag \\
= \gamma v_0 \cos{\phi'(t)}  \label{eq:N5}  \\
-2 m \frac{\text{d}}{\text{d}t}r'\bigg\vert' \frac{\text{d}}{\text{d}t}\theta'\bigg\vert' -mr' \frac{\text{d}^2}{\text{d}t^2}\theta'\bigg\vert' + 2m\omega \frac{\text{d}}{\text{d}t}r'\bigg\vert'-\gamma r' \frac{\text{d}}{\text{d}t}\theta'\bigg\vert' \notag \\ 
= \gamma v_0 \sin{\phi'(t)}\label{eq:N6}
\end{align}

We now place a self-propelled body at initial radial distance $r'(0)=r_A$ and assume $\theta'(0)=0$
without loss of generality. Its initial velocity in the rotating frame is vanishing such that $\frac{\text{d}}{\text{d}t}r'(0)\big\vert'=0$ and $\frac{\text{d}}{\text{d}t}\theta'(0)\big\vert'=0$.
The navigation strategy as determined by the free function ${\phi'(t)}$ is now chosen such that
 the particle starts with an antiradial self-propulsion
$\phi'(0)=-\pi$, moving inwards towards the centre.
During the course of the motion,  $\phi'(t)$ is adjusted such that equation (\ref{eq:N6}) is fulfilled, at any time $t$, hence
$\frac{\text{d}}{\text{d}t}\theta'(t)\big\vert'= \frac{\text{d}^2}{\text{d}t^2}\theta'(t)\big\vert' =0$ so that (\ref{eq:N6}) reads

\begin{equation} \label{eq:N7}
\sin{\phi'(t)} = \frac{2m\omega \frac{\text{d}}{\text{d}t}r'\bigg\vert'}{\gamma v_0}
\end{equation}
Plugging this constraint into Eq. (\ref{eq:N5}) we obtain

\begin{equation} \label{eq:N8}
m \frac{\text{d}^2}{\text{d}t^2}r'\bigg\vert' = m r' \omega^2 -\gamma \frac{\text{d}}{\text{d}t}r'\bigg\vert' - F_0 \sqrt{1-\left(\frac{2 m \omega \frac{\text{d}}{\text{d}t}r'\bigg\vert'}{F_0}\right)^2}
\end{equation}
with the self-propulsion force $F_0=\gamma v_0$ and 
 the initial conditions $r'(t)=r'(0)$ and $\frac{\text{d}}{\text{d}t}r'(0)\big\vert'=0$.  Eq. (\ref{eq:N8}) is physically equivalent to the one-dimensional motion of a particle in the inverted 
parabolic potential $V(r')= -m \omega^2 (r'-R_c)^2/2$ under the nonlinear friction force 
$-f(\frac{\text{d}}{\text{d}t}r'\big\vert') \frac{\text{d}}{\text{d}t}r'\big\vert'/|\frac{\text{d}}{\text{d}t}r'\big\vert'|$ with $f(v)=\gamma |v| - F_0(1-\sqrt{1-4v^2\omega^2/R_c^2})>0$. Consequently the total energy $m\left(\frac{\text{d}}{\text{d}t}r'\big\vert'\right)^2/2+ V(r')$
decreases with time. This analogy shows that the particle will arrive at the origin after a finite time with a finite speed if started with zero velocity at any $ r'<R_c$.

For $ r'_A>R_c$, the centrifugal force exceeds the self-propulsion force. The particle is therefore driven to the outside of the turntable.

\section{Linearly accelerated frame}

\subsection{Equations of motion}

For linear accelerations,  the origin of the accelerated frame is moving on a trajectory
${\vec R}_0(t)$ relative to the origin of an inertial frame. Clearly for linear dependencies
in time,
\begin{equation}\label{eq:L0} 
{\vec R}_0(t)= {\vec R}_0(0) + \vec{V}_0 t,
\end{equation}
we recover the ordinary Galilean
transformation between two inertial frames. For a general ${\vec R}_0(t)$, we get the relation between a trajectory
${\vec r} (t)$ in the inertial frame and that in the accelerated frame, ${\vec r} ' (t)$, by the transformation
\begin{equation} \label{eq:L1}
\vec{r}'(t) = \vec{r}(t)- \vec{R}_0(t)
\end{equation}
and for the orientational degree of freedom clearly
\begin{equation} 
\phi'(t)= \phi(t)
\end{equation} 

The equations of motion in the laboratory frame are
\begin{align} 
m \ddot{\vec{r}} + \gamma (\dot{\vec{r}} - \dot{\vec{R}}_0 (t)) &= \gamma v_0 \hat n + \vec{f} (t)\label{eq:L2}\\
\gamma_R \dot{\phi} &= M+ g(t)\label{eq:L3}
\end{align} 
where we have considered the translational friction proportional to the velocity in the
accelerated frame. The torque balance is not affected by the linear acceleration.
Obviously, the equations of motion are identical to those of a self-propelled particle moving
under the action of an additional external force of
\begin{equation} \label{eq:L2a}
\vec{F}_{\text{ext}}(t)=\gamma \dot{\vec{R}}_0(t)
\end{equation} 
In particular, for a {\it constant\/} relative velocity between the two frames, (\ref{eq:L0}), the external force is constant,
${\vec F}_{ext}(t)=\gamma {\vec V}_0$ and the velocity ${\vec V}_0$ can be interpreted as a drift velocity.
Therefore a Galilean transformation 
is formally equivalent to the action of a constant gravitational force which has been intensely
studied for the overdamped case \cite{ten_Hagen_Nature_Comm_2014,Mazza_EPJE_2019}.
A noise-free circle swimmer will move under the action of a constant force on a curtate or prolate cycloid \cite{trochoids}, a linear swimmer will still swim on a straight line but
along a different direction.
We remark that in the overdamped case $(m=0)$, the correspondence (\ref{eq:L2a}) has further
been exploited for oscillating external forces
in Ref. \cite{Hoffmann_JPCM}.

In the accelerated frame the transformed equations of motion read as
\begin{align} 
m(\frac{\text{d}^2}{\text{d}t^2}\vec{r}'\bigg\vert' - \ddot{\vec{R}}_0 (t)) + \gamma \frac{\text{d}}{\text{d}t}\vec{r}'\bigg\vert' &= \gamma v_0\hat n + \vec{f}(t)\label{eq:L4}\\
\gamma_R \frac{\text{d}}{\text{d}t}\phi'\bigg\vert' &= M+ g(t)\label{eq:L5}
\end{align} 
and look like the equations for a self-propelled particle under the action of the external force
\begin{equation} \label{eq:L6}
{\vec F}_{ext}'(t)= m \ddot{\vec{R}}_0(t)
\end{equation} 
which has been studied for constant acceleration in Ref. \cite{Enculescu_Stark_PRL} in
an external gravitational field.

\subsection{Noise-free trajectories}

The noise-free solutions of Eqs.\ (\ref{eq:L4}) and (\ref{eq:L5}) can be given as

\begin{equation} \label{eq:L7}
\begin{aligned}
&\vec{r}'(t) = \vec{r}'(0) - \dot{\vec{r}}'(0) (e^{-\zeta t}-1)/\zeta \\
&+\zeta^2 v_0^2 \int_0^t \text{d}t' \int_0^{t'} \text{d}t'' e^{-\zeta(t'-t'')}\left(\hat{\mathcal{L}}(e^{i(\phi'(0) + \omega_st')})+ \ddot{\vec{R}}_0 (t')\right) \\
&\cdot \left(\hat{\mathcal{L}}(e^{i(\phi'(0) + \omega_st'')})+ \ddot{\vec{R}}_0 (t'')\right)
\end{aligned}
\end{equation} 
Here the operator $\hat{\mathcal{L}}$ acts on a complex number $Z \in \mathbb{C}$ and produces the vector $\hat{\mathcal{L}}(Z) = (\text{Re}(Z), \text{Im}(Z), 0)^T$, and $\zeta=\gamma/m$.

For a non-inertial frame moving with a constant acceleration $\vec a_0$ along the $x$-axis
relative to the inertial rest frame,  i.e.
${\vec R}_0(t)= {\hat e}_xa_0t^2/2$,
results for the transformed noise-free trajectories
are shown in Figure 6. Under constant acceleration, the trajectory of a linear swimmer transforms into a {\it parabola}.
The noise-free swimming path  of a circle swimmer in the non-inertial frame is a {\it "stretched trochoid"}
in the laboratory frame.

 \begin{figure}
  \centering
  \includegraphics[width=0.45\textwidth]{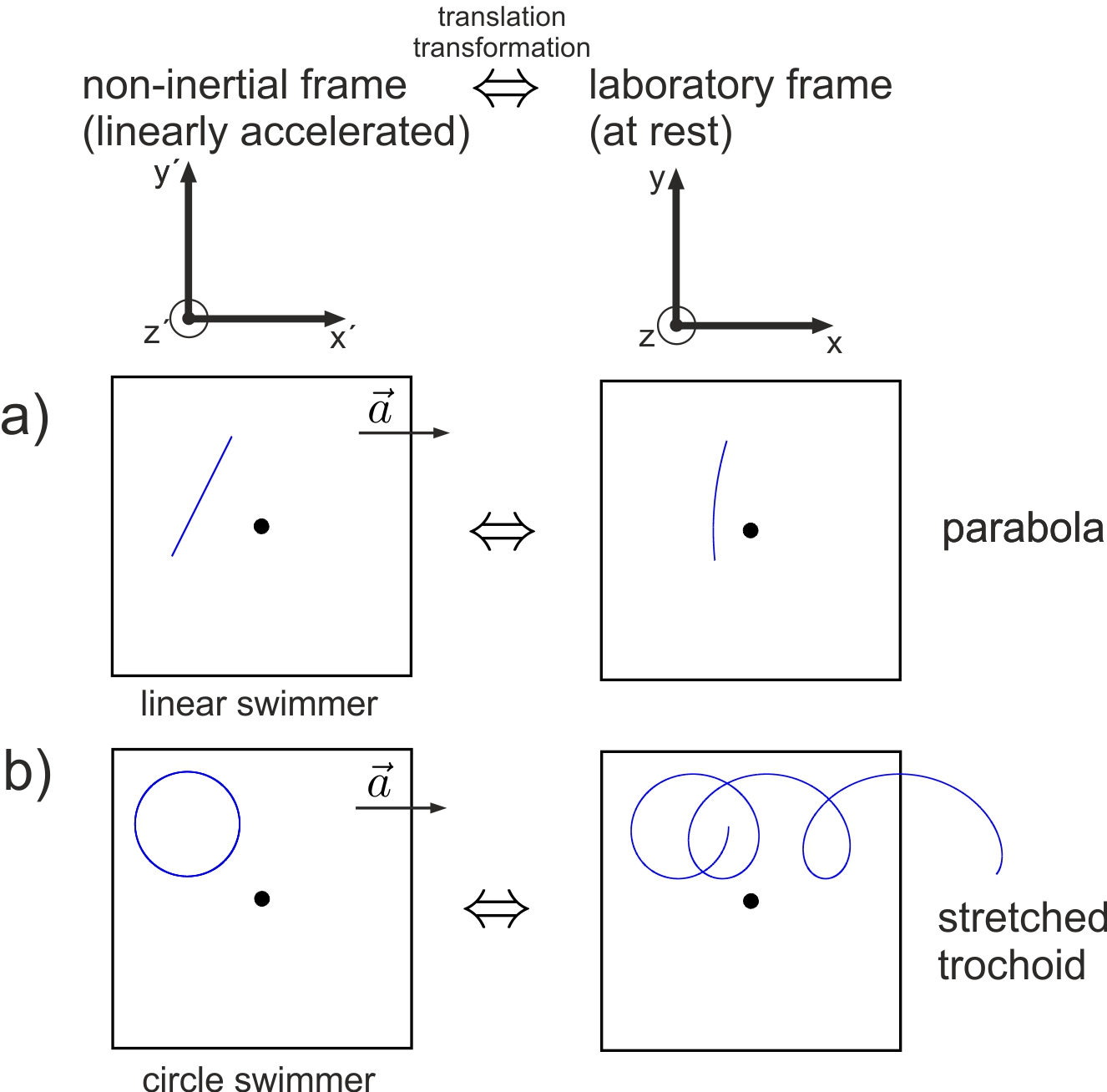}
  \caption{Same as Figure \ref{Fig1} but now for a frame accelerated along the $x$-axis
with a constant linear acceleration $a$. a) The trajectory of a linear swimmer transforms into a parabola.
b)  The swimming path  of a circle swimmer in the non-inertial frame is a "stretched trochoid"
in the laboratory frame.}
\label{Fig6}
\end{figure}

\subsection{Noise averages}

The noise averaged displacement $\overline{ {\vec r}(t) - {\vec r}(0)}$
in the laboratory frame with the prescribed initial condition for
$\dot{\vec r} (0)$ and $\phi(0)$ can be put into relation with  that in the accelerated frame as
\begin{equation} \label{eq:L8}
\overline{\vec{r}(t)-\vec{r}(0)} = \overline{\vec{r}'(t)-\vec{r}'(0)}+ \vec{R}_0(t)-\vec{R}_0(0)
\end{equation} 
where $\overline{ {\vec r}'(t) - {\vec r}'(0)}$ is the noise averaged displacement
in the accelerated frame with the prescribed initial condition for
$\frac{\text{d}}{\text{d}t}\vec r' (0)\bigg\vert'= \dot{\vec r} (0) - \dot{\vec R}_0 (0) $ and $\phi'(0)=\phi(0)$.
For $\overline{ {\vec r}'(t) - {\vec r}'(0)}$ we obtain

\begin{align}
&\overline{\vec{r}'(t)-\vec{r}'(0)} = (1-e^{-\zeta t}) (\dot{\vec{r}}(0)-\dot{\vec{R}}_0(0))\zeta + \zeta v_{0}\hat{\mathcal{L}} \left(\frac{e^{i \phi(0)}}{\zeta} \right) \notag \\
&+\zeta v_{0}\hat{\mathcal{L}} \left( - D_r + i \omega_s \left[\frac{e^{(-D_r + i \omega_s)t}-1}{-D_r} + i \omega_s + \frac{(e^{-\zeta t}-1)}{\zeta}\right]\right)\label{eq:L9}
\end{align} 
In the overdamped limit ($m=0$), this reduces to Eq.\ (\ref{eq:18}).

\begin{figure}
  \centering
  \includegraphics[width=0.50\textwidth]{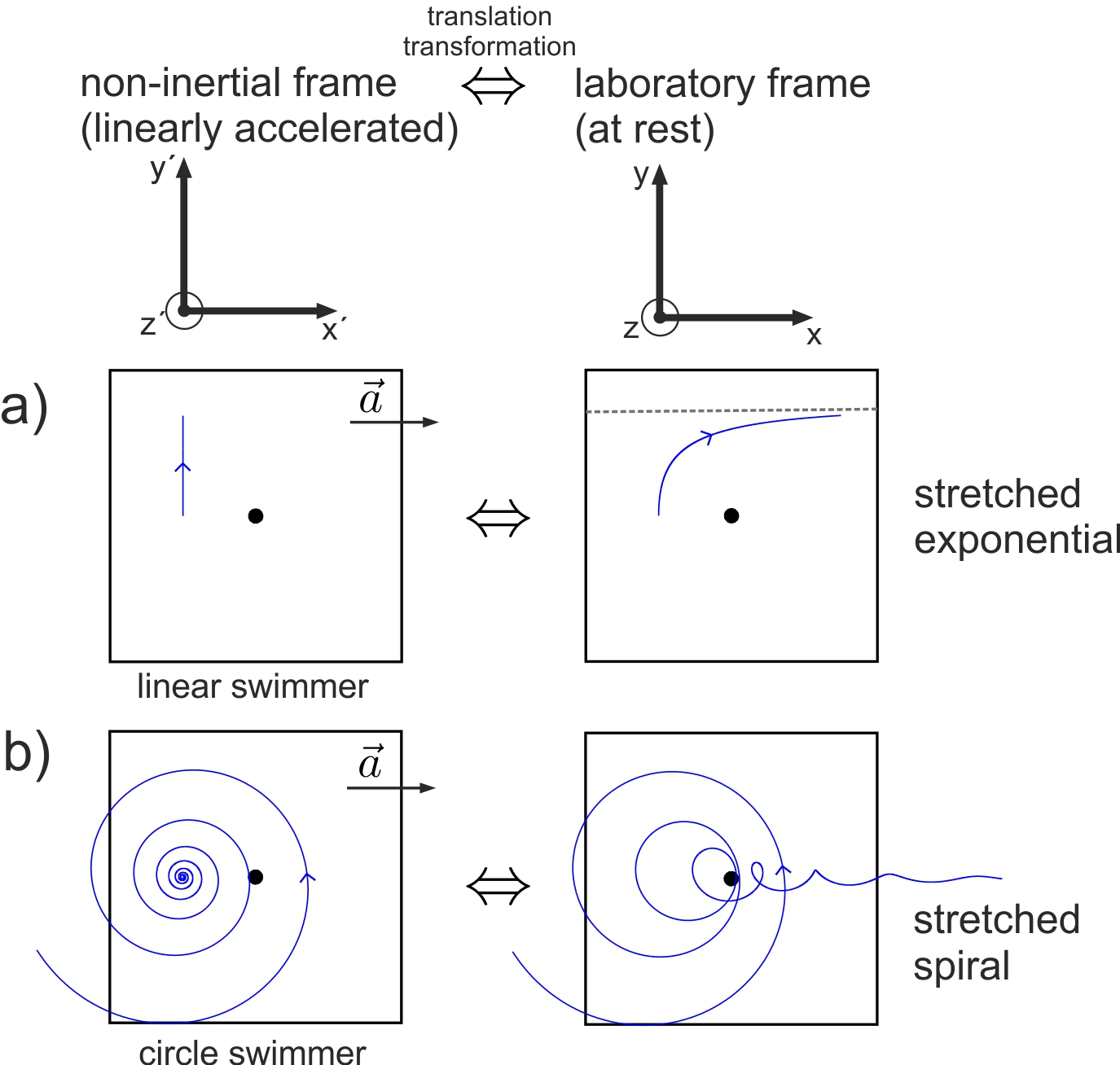}
  \caption{Same as Figure \ref{Fig2} but now for a frame accelerated along the $x$-axis
with a constant linear acceleration $a$. The mean displacements are shown for the overdamped case
for a) linear swimmers and b) circle swimmers. The straight linear segment transforms into a stretched exponential
if the initial orientation is perpendicular to the acceleration $\vec{a}$. The {\it spira mirabilis\/} of a circle swimmer
transforms into a stretched spiral.}
\label{Fig7}
\end{figure}

The transformed averaged displacements are shown in Figure 7 in the overdamped case.
The straight segment for a linear swimmer (\ref{eq:18}) transforms into the curve

\begin{equation} \label{eq:L12}
y(x) = \sin{(\phi'(0))} \frac{v_0}{D_r} (1-e^{-D_r \sqrt{\frac{2}{a}(x-y/\tan{\phi'(0)})}})
\end{equation} 
which for the special case of an initial propulsion perpendicular to the acceleration, $\phi'(0)=\pi/2$
is a {\it stretched exponential\/} function

\begin{equation} \label{eq:L13}
y(x) = \frac{v_0}{D_r} (1-e^{-D_r\sqrt{\frac{2x}{a}}})
\end{equation} 
For circle swimmers, the logarithmic spiral transforms into a {\it stretched spiral}
reminiscent to an unreeling helix, see Figure \ref{Fig7}b.
 
Finally, the  noise averaged mean-square displacement $\overline{ (\vec{r}(t) - \vec{r}(0))^2}$
in the laboratory frame with the prescribed initial condition for
$\dot{\vec r} (0)$ and $\phi(0)$ can be expressed in terms of

\begin{align}
 &\overline{(\vec{r}(t)-\vec{r}(0))^2} =  \overline{(\vec{r}'(t)-\vec{r}'(0))^2} \notag \\
 &+ 2 \overline{(\vec{r}'(t)-\vec{r}'(0))} \cdot (\vec{R}_0(t)-\vec{R}_0(0)) +(\vec{R}_0(t)-\vec{R}_0(0))^2\label{eq:L10}
\end{align} 
where $\overline{( {\vec r}'(t) - {\vec r}'(0))^2}$
is the noise averaged displacement
in the accelerated frame with the prescribed initial condition for
$\frac{\text{d}}{\text{d}t}{\vec r}' (0)\big\vert'= \frac{\text{d}}{\text{d}t}\vec{r} (0)\big\vert' - \dot{\vec R}_0 (0) $ and $\phi'(0)=\phi(0)$
and for $\frac{\text{d}}{\text{d}t}\vec{r}' (0)\big\vert'= \dot{\vec r} (0) - \dot{\vec R}_0 (0) $. We obtain

\begin{widetext}
\begin{align}
&\overline{(\vec{r}'(t)-\vec{r}'(0))^2} = \frac{1}{\zeta^2} \left(\dot{\vec{r}}(0)-\dot{\vec{R}}_0(0)\right)^2 \cdot{(1-e^{-\zeta t})^2} \notag \\
&+2 v_0 (1-e^{-\zeta t}) \left(\dot{\vec{r}}(0)-\dot{\vec{R}}_0(0)\right) \cdot \hat{\mathcal{L}} \left( \frac{e^{i\phi(0)}}{\zeta - D_r + i \omega_s} \left[ \frac{e^{(-D_r+i\omega_s)t} -1}{-D_r+i\omega_s}+\frac{e^{-\zeta t}-1}{\gamma}\right]\right) +2 Dt + 2 \frac{D}{\zeta} \left(e^{-\zeta t} -1-\frac{(1-e^{-\zeta t})^2}{2}\right) \notag \\
&+\text{Re}\left\{ \frac{2 v_0^2}{\zeta(D_r-i\omega_s)}\left[\frac{t}{\zeta} + \frac{1-e^{-(\zeta+D_r-i\omega_s)t}}{(\zeta-D_r)^2-(i\omega_s)^2}  + \frac{(e^{-\zeta t}-1)}{\zeta^2} +(D_r-i\omega_s)\frac{(1-e^{-\zeta t})^2}{2\zeta^2 (\zeta-D_r+i \omega)} + \frac{e^{(-D_r + i\omega_s)t}-1}{(D_r - i\omega_s)(\zeta-D_r+i\omega_s)}\right]\right\}\label{eq:L11}
\end{align} 
\end{widetext}
which was previously obtained in the special case of linear
swimmers ($\omega_s=0$) in Ref.\ \cite{Zippelius}. Again, in the overdamped limit ($\zeta\to\infty$), the expression (\ref{eq:L11}) reduces to (\ref{eq:20}).

For a constantly accelerated frame  ${\vec R}_0(t)= {\hat e}_xa_0t^2/2$, the mean-square displacement
is "superballistic" and  will
grow with a power law for long times

\begin{equation*}
\overline{ ({\vec r}(t) - {\vec r}(0))^2} \simeq  a^2 t^4 / 4
\end{equation*}
resulting from the acceleration. In this case the long-time self-diffusion coefficient $D_L$ (see (\ref{eq:23})) does not exist
(it is rather diverging), but the long-time self-diffusion coefficient  $D_L'$ exists in the non-inertial frame.
An analytical expression for $D_L'$ was given in Ref.\ \cite{Ldov}.

\subsection{Navigation strategy}
For a particle which can adjust its orientation, an optimal swimming strategy to navigate somewhere can be obtained
by counter-aligning the particle orientation against the inertial force
${\vec n}'(t) = -\ddot{{\vec R}}_0(t)/ |\ddot{\vec{R}}_0(t)|$.
If the condition
\begin{equation*}
\gamma v_0 > m |\ddot{\vec{R}}_0(t)|
\end{equation*}
is fulfilled for any time $t$,
the self-propulsion
force will be always larger than the inertial force such that the additional
freedom in orientation can be used to navigate to  an arbitrary target point.

\section{Experimental verification of the predictions} 
The best realization of our model equations can be found for active granulates
\cite{Ramaswamy,Dauchot_Frey_PRL_2013,Poeschel,Deblais,Tsimring,Chate,Patterson,Dauchot2,Mayya1,Mayya2}.
Typically these are hoppers with
a Janus-like body or with tilted legs. In order to achieve self-propulsion, these macroscopic
bodies are either placed on a vibrating table or are equipped with an internal vibration motor
("hexbugs") \cite{Dauchot_hexbugs}. In a rest frame, it has been shown that
the dynamics of these hoppers is  well described by
active Brownian motion with inertia \cite{Walsh,Dauchot_Brownian,Ldov}. Since they are macroscopic, inertia is relevant,
the fluctuations can be fitted to Brownian forces, and imperfections in the particle symmetry
will make them circling. Therefore, "hexbugs" on a turntable or granulates on a vibrating rotating plate are a
direct realization of the phenomena discussed in II.B. A turntable which is the standard tool to demonstrate
fictitious forces in classical physics \cite{Levine1978,Bligh1982,Gersten1992,Ehrlich1995} just needs to be loaded with a "hexbug" \cite{ScholzPrivate}.
Then the deterministic trajectories can be watched directly and noise-averages are accessible by
averaging over different trajectories.

Navigation strategies as discussed in II.C can be implemented by a feedback
coupling to turn the particle orientations. In a similar context this has been done
to implement the motion of mini-robots \cite{Wehr,Volpe2}.

Linear accelerations discussed in III can also in principle be realized by studying self-propelled granulates
on a horizontally oscillating plate.
The special situation of constant linear acceleration is obtained by tilting
the substrate \cite{ten_Hagen_Nature_Comm_2014} since this includes the action of gravity.
Finally, combinations of translational and rotational accelerations
 can be realized by a tilted turntable \cite{Romer1981}.

In the overdamped limit, there are many standard examples of low-Reynolds number microswimmers on substrates.
These can either be synthetical Janus-colloids or microorganism like bacteria and sperm. Both linear \cite{Houwse} and circle swimmers
\cite{Ecoli,ten_Hagen_Nature_Comm_2014,Ebbens} have been studied in the rest frame. The overdamped version of our equations
considered in II.A are realized in a rotating container filled with
fluid \cite{Liebchen_navigation} which is rotating with a constant angular velocity $\omega$.
Then the stick boundary condition near a substrate surface will enforce a
solvent velocity flow field which is approximately given by $\vec{u} (\vec{r})= \vec{\omega} \times \vec{r}$.
This velocity field has a constant vorticity $\nabla \times {\vec u} ({\vec r}) = 2 \vec{\omega}$.
In the laboratory frame, an active particle is therefore advected by the flow field and simultaneously rotated
by the local shear rate $\dot{\gamma} /2$
\cite{JanDhontbook_Dynamics_of_colloids,tenHagen_Wittkowski_PRE_2011,Tarama_2014}. These are exactly
the overdamped equations of motion considered in II.A. Linear accelerations can be realized by appropriate
time-dependent external forces such as electric fields, magnetic field gradients or time-dependent gravity
(as obtained from  turning the substrate horizontally with an appropriate time-dependent rotation speed).
Macroscopic swimmers embedded in a rotating fluid will exhibit inertial effects on top of the frictional ones. One example are waterlily beetles moving near the two-dimensional  water-air interface
\cite{beetles}.

In a more general sense, other realizations are conceivable: First, our
planet is rotating and therefore a non-inertial frame.
Airplanes and flying birds are self-propelled objects and therefore the combinations of centrifugal, Coriolis
and self-propulsion forces should play a major role for their dynamics. Second, dust particles
in plasmas ("complex plasmas")
can be made active \cite{Bartnick} and exhibit underdamped dynamics due to the presence of the neutral gas
\cite{Ivlev_Morfill}. Confining an active complex plasma  between two rotating electrodes
\cite{Bonitz,Hartmann1,Hartmann2} would correspond to
another of the equations of motion studied in II.B. Last, the actual motion of humans on turntables and
carousels is a third example where our equations should apply.

\section{Conclusions} 

In summary, while the frictional motion of passive objects are well-studied over decades
\cite{Weltner1979,Sokirko1994,Weckesser1997,Agha2015} and
serve as a simple demonstration of the action of fictitious forces (such as the Coriolis forces and the
centrifugal force)
\cite{Levine1978,Bligh1982,Gersten1992,Ehrlich1995},
we have upgraded the dynamics here by including self-propulsion and Brownian noise
in the non-inertial frame (albeit using Stokes friction rather than solid-on-solid friction). We thereby
link the classic problem of a body on a turntable
to the expanding field of active matter. 

In the overdamped case of vanishing particle mass where inertial effects are absent,
most of the physics can be obtained by a pure coordinate transformation from the rest into the moving frame.
Still this results in new epicyclic swimming paths in the rest frame mathematically
described by epitrochoids, hypotrochoids and stretched trochoids.
For non-zero particle mass there are additional centrifugal and Coriolis forces at work which
lead to swimming paths on  logarithmic spirals outwards the rotation centre.
For a particle initially at rest in a rotating non-inertial frame,
a swimming strategy to stay close to the origin can be given if the initial distance to
the origin is smaller than a critical radius $R_c$. The results are verifiable in various experimental set-ups.

Future studies should be performed along the following lines:
first,  more general situations should be treated numerically and analytically.
In particular the situation of finite orientational relaxation time ($J>0$) is promising \cite{Ldov}.
Second,  anisotropic particles (such as rod-like swimmers) should be considered in which case the
translational fluctuations are anisotropic \cite{Kuemmel_et_al_PRL_2013} and will therefore
be different in the rotating and rest frame. In particular particles with an anisotropic mass distribution are expected to exhibit similiar effects as bottom heavy swimmers under gravity \cite{Wolff}.
Next, swimming in full three spatial dimensions is more complicated but  relevant \cite{Wysocki_Gompper}.
Fourth, there is an analogy between the Lorentz force acting on a charged particle in the rest
frame and the Coriolis force acting on uncharged particles in the rotating frame \cite{Burns,Bonitz,Sandoval}.
Therefore our methods will be profitable also to study
charged swimmers in a magnetic field. However, the actual magnetic fields required to see an
effect of a bent swimmer trajectory needs to be immense, even for highly charged dusty plasmas \cite{Bonitz}.

Finally an ensemble of many particles should be considered in a rotating frame. In a rest frame,
collections of circular swimming particles have been studied in various situations \cite{Kuemmel_et_al_PRL_2013, Ref30_von_Klapp,Ebbens,Ref32_von_Klapp,Ref33_von_Klapp,Kurzthaler,Soudeh,Ref36_von_Klapp,Ref37_von_Klapp,Ref38_von_Klapp,Ref39_von_Klapp,Ref40_von_Klapp,Levis,Ref42_von_Klapp}
and many linear swimmers with inertia have been more recently explored \cite{Mandal}. An open
question is whether in a rotating frame there is a gradient of kinetic temperature
maintained and how this affects motility-induced phase separation \cite{Fily,Tailleur}.

\acknowledgments
I thank Soudeh Jahanshahi, Christian Scholz, Alexander Sprenger, Andreas M. Menzel, Alexei V. Ivlev,
Frederik Hauke and Benno Liebchen for helpful discussions and
gratefully acknowledge support by the Deutsche
Forschungsgemeinschaft (DFG) through grant
LO 418/23-1.


%

\end{document}